\documentclass[twocolumn,superscriptaddress,aps,prb,showpacs,showkeys]{revtex4-1}
\usepackage[T1]{fontenc}
\usepackage[latin9]{inputenc}
\usepackage{color}
\usepackage{float}
\usepackage{amsmath}
\usepackage{graphicx}
\usepackage{amssymb}
\usepackage{esint}

\begin{document}

\title{Dynamic quantum Kerr effect in circuit quantum electrodynamics}

\author{Yi Yin}
\affiliation{Department of Physics, University of California, Santa Barbara, CA 93106-9530, USA}
\author{H. Wang}
\thanks{Present address: Department of Physics, Zhejiang University, Hangzhou 310027, China.}
\affiliation{Department of Physics, University of California, Santa Barbara, CA 93106-9530, USA}
\author{M. Mariantoni}
\affiliation{Department of Physics, University of California, Santa Barbara, CA 93106-9530, USA}
\affiliation{California NanoSystems Institute, University of California, Santa Barbara, CA 93106-9530, USA}
\author{Radoslaw C. Bialczak}
\affiliation{Department of Physics, University of California, Santa Barbara, CA 93106-9530, USA}
\author{R. Barends}
\affiliation{Department of Physics, University of California, Santa Barbara, CA 93106-9530, USA}
\author{Y. Chen}
\affiliation{Department of Physics, University of California, Santa Barbara, CA 93106-9530, USA}
\author{M. Lenander}
\affiliation{Department of Physics, University of California, Santa Barbara, CA 93106-9530, USA}
\author{Erik Lucero}
\affiliation{Department of Physics, University of California, Santa Barbara, CA 93106-9530, USA}
\author{M. Neeley}
\thanks{Present address: Lincoln Laboratory, Massachusetts Institute of Technology, 244 Wood Street, Lexington, MA 02420-9108, USA.}
\affiliation{Department of Physics, University of California, Santa Barbara, CA 93106-9530, USA}
\author{A. D. O'Connell}
\affiliation{Department of Physics, University of California, Santa Barbara, CA 93106-9530, USA}
\author{D. Sank}
\affiliation{Department of Physics, University of California, Santa Barbara, CA 93106-9530, USA}
\author{M. Weides}
\thanks{Present address: National Institute of Standards and Technology, Boulder, CO 80305, USA.}
\affiliation{Department of Physics, University of California, Santa Barbara, CA 93106-9530, USA}
\author{J. Wenner}
\affiliation{Department of Physics, University of California, Santa Barbara, CA 93106-9530, USA}
\author{T. Yamamoto}
\affiliation{Department of Physics, University of California, Santa Barbara, CA 93106-9530, USA}
\affiliation{Green Innovation Research Laboratories, NEC Corporation, Tsukuba, Ibaraki 305-8501, Japan}
\author{J. Zhao}
\affiliation{Department of Physics, University of California, Santa Barbara, CA 93106-9530, USA}
\author{A. N. Cleland}
\affiliation{Department of Physics, University of California, Santa Barbara, CA 93106-9530, USA}
\affiliation{California NanoSystems Institute, University of California, Santa Barbara, CA 93106-9530, USA}
\author{John M. Martinis}
\email{martinis@physics.ucsb.edu}
\affiliation{Department of Physics, University of California, Santa Barbara, CA 93106-9530, USA}
\affiliation{California NanoSystems Institute, University of California, Santa Barbara, CA 93106-9530, USA}


\begin{abstract}
A superconducting qubit coupled to a microwave resonator provides a controllable system that enables fundamental studies
of light-matter interactions. In the dispersive regime, photons in the resonator exhibit
induced frequency and phase shifts which are revealed in the resonator transmission spectrum measured with fixed qubit-resonator detuning.
In this \emph{static} detuning scheme, the phase shift is measured in the far-detuned, linear
dispersion regime to avoid measurement-induced demolition of the qubit quantum state. Here we explore
the qubit-resonator dispersive interaction over a much broader range of detunings, by using
a \emph{dynamic} procedure where the qubit transition is driven adiabatically. We use resonator Wigner tomography to monitor the interaction, revealing
exotic non-linear effects on different photon states, e.g., Fock states, coherent states, and Schr\"{o}dinger cat states, thereby demonstrating
a quantum Kerr effect in the dynamic framework.
\end{abstract}

\pacs{42.50.Pq, 42.50.-p, 85.25.-j, 03.65.Wj}

\maketitle

\section{INTRODUCTION}
A major focus of quantum optics is the study of atom-photon interactions
at the microscopic level.\cite{QuantumOptics}
Atomic and solid-state cavity quantum electrodynamics (QED) systems,\cite{CavityQEDMabuchi,OpCavity,QDotETH,CavityQEDBook} with strongly enhanced
coupling strength between confined atoms and photons,
have been developed to pursue this goal. The superconducting version, circuit QED,
provides a test bed for quantum microwave photons\cite{Horizons} interacting with superconducting qubits.\cite{ScQubitReview,QuaEnginneering,PhysToday}
By placing a one-dimensional transmission line resonator in close proximity
to a superconducting qubit,\cite{Yale1,max1,SimmondsBusNature,IBM,NTT,neccQED,ChiorescuNat04}
the electromagnetic fields in the resonator interact strongly with the qubit,
forming a composite system with a modified energy spectrum (see Fig. 1(b)).
The resonator photon energy is effectively dispersed by the qubit, with a dispersion
strength that depends strongly on the qubit-resonator frequency detuning. In
previous experiments, the frequency and phase shift of transmitted photons
were measured with a fixed qubit-resonator detuning, which we term the \emph{static} scheme.
Although the photon dispersion
is well understood in this regime and provides a direct qubit
readout scheme,\cite{wallraffPRL05} it is
restricted to measurements with the qubit far detuned from the resonator to avoid
measurement-induced demolition of the qubit quantum state.\cite{SchusterNat}

In this paper, we introduce an alternative
\emph{dynamic} scheme for measuring qubit-photon interactions. As we adiabatically
change the qubit frequency, we measure
the accumulated phase shift of the photons while minimizing energetic interactions with the qubit.
The detuning can range from large detuning corresponding to the linear
response regime, to strong nonlinear dispersion found near and at resonance.
The photonic response is measured with Wigner tomography, which enables
a complete map of the photon state after a dispersive interaction
with the qubit.\cite{max2} We further
investigate the response of completely nonclassical photon states, beyond the scope of prior static measurements,
which have been restricted to measurements of classical coherent states. Measurements of the nonlinear
dispersive response of superpositions of Fock states and Schr\"odinger cat states exhibit an excellent
correspondence with expected dynamic response and demonstrate a quantum version of the Kerr effect. Classically, the Kerr effect,
a nonlinear response of macroscopic materials is where the refractive index of a medium changes when varying the electromagnetic
field strength. In this quantum version, a single qubit medium induces Fock-number dependent phase shifts to
quantum state photons. Although the near-resonant nonlinear response of coupled atom-photon systems was previously explored
using two-tone\cite{ladder,Deppe} and high power spectroscopy,\cite{nonlinear,Reed,OpCavity,cavityRabi}
our approach delineates the practical aspects associated with the strong non-linear phase shift of photons
in the time domain. In the future, it should be possible to exploit the large photon phase shift resulting from small
qubit-resonator detunings to improve the quantum non-demolition qubit readout with higher fidelity.

\section{EXPERIMENT AND RESULTS}

We use a half-wavelength superconducting coplanar waveguide (CPW) resonator coupled
to a superconducting Josephson phase qubit (Fig. 1(a)). The resonator has a fixed
frequency $\omega_{r}/2\pi\sim6.32\ \text{GHz}$
and an energy relaxation time $T_{1}\sim2\ \mu\text{s}$. The flux-biased
phase qubit is a strongly nonlinear electrical quantum circuit, with a tunable
$|g\rangle\leftrightarrow|e\rangle$ transition angular frequency $\omega_q$ and $|e\rangle\leftrightarrow|f\rangle$
transition frequency $\sim200\ \text{MHz}$ smaller ($|g\rangle$, $|e\rangle$ and $|f\rangle$
are the ground, excited, and third level qubit states).\cite{John} The qubit angular frequency $\omega_q$ can be tuned
rapidly using quasi-DC $\sigma_{z}$ pulses on the inductively-coupled qubit flux bias
line; qubit energy level transitions are driven by $\pi$-pulse ($\sigma_{x,y}$) resonant microwave signals applied to the same line.\cite{ErikPRL,ErikPRA} The qubit state can be measured destructively using a single-shot readout involving short flux pulses followed by measurement using an on-chip superconducting quantum interference device (SQUID).\cite{MatthewPRB} The qubit has a relaxation time $T_{1}\sim450$ ns and a phase decoherence time $T_{2}\sim150$ ns. The qubit and resonator are capacitively coupled with a coupling strength $g/2\pi\sim9\ \text{MHz}$ (i.e., the qubit-resonator gap from spectroscopy is $\sim18\ \text{MHz}$ at resonance).

\begin{figure}[H]
\begin{center}
\includegraphics[width=1.0\columnwidth,clip]{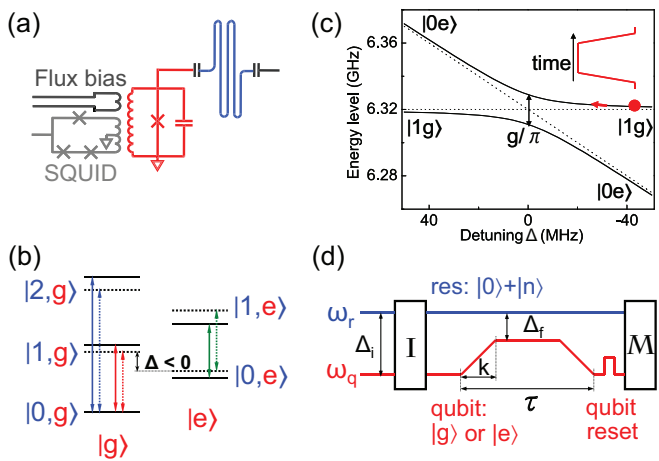}
\caption{~\label{Fig1}~(Color online)
(a) Circuit diagram for the coupled qubit-resonator system. The superconducting
phase qubit (red) has a ground to excited state transition frequency $\omega_q/2\pi\sim6\ \text{GHz}$.
The flux bias line (black) is coupled to the qubit through a mutual inductance.
A three-junction superconducting quantum interference device (SQUID) (grey)
is optimized to read out the qubit state. The CPW resonator (blue)
has a constant resonant frequency $\omega_r/2\pi\sim6.32\ \text{GHz}$. The resonator is weakly coupled
to both the qubit and external microwave source
by small capacitors. (b)~Energy levels
of the coupled system. The solid (dashed) lines
denote the energy levels with (without) the qubit-resonator interaction,
with $|n,q\rangle$ representing the photon number $|n\rangle$ and qubit state $|q\rangle$.
The red and blue arrows represent one- and
two-photon transitions with the qubit ground state. The green arrows
represent one-photon transitions with the qubit excited state. (c) Plot of the $|1,g\rangle$ and $|0,e\rangle$ energies
as a function of detuning $\Delta = \omega_q-\omega_r$, calculated from the Jaynes-Cummings model
using our experimental parameters. In an adiabatic process, the system
initially at $|1,g\rangle$ (red dot) follows the path indicated by
the red arrow. The resonant avoided-level crossing is $g/\pi\sim18\ \text{MHz}$. (d) Schematic of the experimental sequence, describing system initialization (labeled by ``I''), adiabatically tuning and detuning the qubit, and finally performing Wigner tomography measurement (labeled by ``M'')
of the final photon state. A reset pulse\cite{MatteoScience} to ground is applied to the qubit if it was initially prepared in the excited state, to
allow the qubit to perform the measurement. The
dispersive part of the pulse is also shown schematically in (c). }
\end{center}
\end{figure}

As shown for the specific case of $|0,e\rangle\leftrightarrow|1,g\rangle$ in Fig. 1(c) [where states are denoted by $|n,q\rangle$ for resonator state $|n\rangle$ and qubit state $|q\rangle$ (Fig. 1(b))], the dynamic dispersion scheme involves an adiabatic procedure\cite{photonQND} to vary the detuning $\Delta = \omega_q-\omega_r$ and thus the eigenenergies of the qubit-resonator system. We initialize the coupled system in the far-detuned regime, where the dispersive interaction is extremely weak. We then smoothly adjust the qubit frequency as a function of time $t$, so that the coupled system adiabatically follows its instantaneous eigenstates with eigenenergies $E(t)$. The energy of a single photon (relative to the vacuum) is thus dispersed to $e_{p}(t) = E_{|1,g\rangle}(t)-E_{|0,g\rangle}(t)$, and the dynamic phase\cite{berry} accumulates as $\theta(\tau)=-\frac{1}{\hbar}\int_{0}^{\tau}[e_{p}(t)-e_{p}(0)]dt$. Although the system is returned to its initial settings at the end of the interaction, an accumulated phase shift is acquired even though the net energy exchange is zero. A Berry's phase component is ruled out because the parameter space is one-dimensional. We note that the qubit-resonator coupling strength is to all practical purposes constant in our system and the detuning is time-dependent, in contrast to the time-varying coupling and constant detuning displayed by Rydberg atoms passing through a microwave cavity.\cite{cavityQED,cavityphotonQND,cavityLamb}

We first investigated the phase shift of a one-photon Fock state due to its interaction with a
qubit in its ground state. The resonator was initialized in the Fock state superposition $|0\rangle+|1\rangle$,
where $|0\rangle$ serves as a reference state.\cite{qudit} The detuning was set to
$\Delta_{i}/2\pi=-250\ \text{MHz}$, giving a small photonic component $g/\Delta_{i}\sim4\%$.
We then applied a trapezoidal tuning pulse to the qubit, with linear ramp rate $k=5\ \text{MHz/\text{ns}}$
and interaction detuning $\Delta_{f}$ for an adjustable time $\tau$ (Fig. 1(d)). At the end of the tuning pulse, the qubit was effectively
decoupled from the resonator and in its ground state, with a
small error due to the non-ideal adiabatic process. The Wigner tomogram of the final resonator
state was reconstructed by displacing the resonator state with a coherent microwave pulse (with complex amplitude
$\alpha$) and performing a qubit-resonator swap, from which we extract
the photon state probability $P_{n}(\alpha)$.\cite{max2,haohua2} With sufficient sampling points $\alpha$
in the resonator phase space, we calculate the resonator density matrix and reconstruct the quasi-probability $W(\alpha)$.\cite{haohua2}

We plot the resulting Wigner functions for detuning $\Delta_{f}/2\pi=-57\ \text{MHz}$ and five
different durations $\tau$ in the top row of Fig. 2(a). Negative quasi-probabilities are a signature
of the expected non-classicality of the superposed Fock states.
The diagrams are not rotationally symmetric, encoding the relative
phase shift between the $|0\rangle$ and $|1\rangle$ states.  As the dispersion duration $\tau$ increases, the
Wigner function rotates counter-clockwise about the origin. The rotation is due to an overall phase shift, induced
by the qubit state during the dispersive interaction. We plot resonator
density matrices in the
Fock basis in the bottom row of Fig. 2(a). In each density matrix, the population
probabilities $\rho_{00}$ and $\rho_{11}$ are nearly identical,
but energy relaxation becomes more pronounced as $\tau$
increases. For $\tau =0\ \text{ns}$, the off-diagonal terms $\rho_{10}$
and $\rho_{01}$ are represented by horizontal arrows, indicating no phase shift as expected. As
$\tau$ increases, the arrow representing $\rho_{10}$ rotates with the
same angle as that of the overall Wigner function, exhibiting an accumulated
non-zero phase shift; the time dependence of this phase rotation angle is shown in Fig. 2(b). The constant rotation rate represents a phase rotation speed $r_0=d\theta/d\tau=3.1\text{\ensuremath{\pi} \text{rad}/\ensuremath{\mu}\text{s}}$ and is equal to the expected frequency shift of the one-photon
state $|1\rangle$ at the detuning $\Delta_{f}$.

\begin{figure}[H]
\begin{center}
\includegraphics[width=1.0\columnwidth,clip]{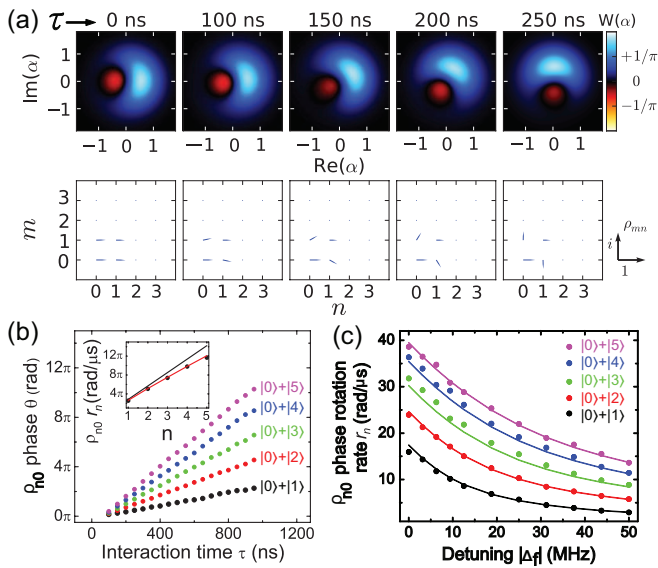}
\caption{~\label{Fig2}~(Color online)
Wigner tomography and phase rotation measurement for the initial resonator
state $|0\rangle+|n\rangle$. (a) Reconstructed
Wigner functions for $|0\rangle+|1\rangle$ after a dispersive interaction
with a ground state qubit at $\Delta_{f}/2\pi=-57\ \text{MHz}$ for
five different value of $\tau$ . The corresponding density matrices are shown
in the bottom row. Sampling points\cite{haohua2} for $W(\alpha)$ comprised
60 points around two concentric circles with radii
$|\alpha|=1.10\ \text{and}\ 1.45$. (b) Rotation angle of $\rho_{n0}$
as a function of $\tau$, measured from its value at
$\tau=100\ \text{ns}$. (Inset) The $n$-dependent phase rotation rate is obtained from a linear fit to
the data in (b). The results from both the linear approximation (black
line) and the exact diagonalization (red line) of the Jaynes-Cummings
Hamiltonian are displayed without any fit parameters. (c) Phase rotation speed of $\rho_{n0}$ as a function of both $\Delta_{f}$
and $n$. Solid lines are theoretical predictions from the exact
diagonalization of the Jaynes-Cummings Hamiltonian.}
\end{center}
\end{figure}

We next investigated higher number photon states by initializing the
resonator in $|0\rangle+|n\rangle$ with $n$ from $2$
to 5. Following the same pulse sequence, the density matrices
and the Wigner functions of the final photon states were measured.
Figure 2(b) shows the phase rotation angles of the $\rho_{n0}$ element of the density matrix
versus $\tau$. Linear fits allow us to extract the $n$-dependent phase rotation speed. In the
far-detuned regime, the angular frequency shift scales as expected with photon number
as $ng^{2}/|\Delta_{f}|$ (cf. Eq. 3 in Appendix B). This
linear approximation (black line, Fig. 2(b) inset) deviates from
experimental data as $n$ increases, because the dressed photon
energies become anharmonic; a rigorous solution of the Jaynes-Cummings
model (cf. Eq. 1 in Appendix B) is in good agreement with the data (red line, Fig. 2(b) inset). The phase rotation rate as a
function of detuning $\Delta_f$ (Fig. 2(c)) is also in excellent agreement with the full Jaynes-Cummings
model, extending the nonlinear response measurements to zero detuning. This rate corresponds to the refractive
index of macroscopic materials, where a single qubit, as an effective electric dipole, is the index medium which interacts
with the electric field of quantum state photons. The Fock state dependent phase shifts imply a photon number dependent
change in the refractive index of the medium, demonstrating a quantum Kerr effect in the dynamic framework.
Although the energy levels of the qubit-resonator system can be measured using
spectroscopy,\cite{WallraffLamb} here the nonlinear phase shift in the strong dispersive regime is
directly and quantitatively measured in the time domain in a circuit QED system.

\begin{figure}[H]
\begin{center}
\includegraphics[width=0.9\columnwidth,clip]{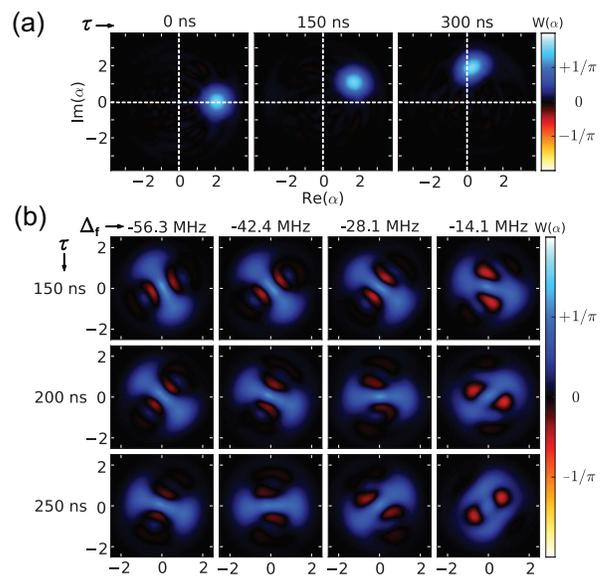}
\caption{~\label{Fig3}~(Color online)
(a) Wigner functions and density matrices for a coherent state ($\alpha\sim2.0$)
interacting with a ground state qubit at three different durations $\tau$ for detuning $\Delta_{f}/2\pi=-50\ \text{MHz}$. Axes
at the origin are indicated. The
center of the coherent-state Wigner function rotates counter-clockwise about the origin with time. (b) Wigner functions for a Schr\"odinger cat state $|\alpha=\sqrt{2}i\rangle+|\alpha=-\sqrt{2}i\rangle$
after different interaction times with a ground state qubit. A
fixed detuning $\Delta_f$ from $-56.3\ \text{MHz}$ to $-14.1\ \text{MHz}$ was used for each column. As the duration
$\tau$ increases from $150\ \text{ns}$ to $250\ \text{ns}$, the Wigner function rotates counter-clockwise, with rotation rate increasing as the absolute value of detuning $\Delta_{f}$ decreases. For small detunings, enhanced distortions appear in both the coherent peaks and the interference fringes.}
\end{center}
\end{figure}

In a standard resonator transmission measurement, a coherent state $|\alpha\rangle$
is injected in the resonator by a classical microwave pulse. Here
we prepared a coherent resonator photon state with complex amplitude $\alpha\sim2.0$ and applied
the dynamic scheme to measure the phase shift induced by interaction
with a ground-state qubit. Figure 3(a) presents the measurement after a trapezoid
pulse at $\Delta_{f}/2\pi=-50\ \text{MHz}$. The Wigner function of
the initial state ($\tau=0\ \text{ns}$) shows a coherent peak centered
at $\alpha$ and a symmetric noise of quantum uncertainty. As $\tau$
increases, the coherent peak rotates counter-clockwise about the origin,
with only a slight and gradual increase in distortion. For
each Fock constituent, the linear approximation gives a phase shift rate
$r_{n}\sim nr_{0}=ng^{2}/|\Delta_{f}|$. The coherent state is thus
uniformly transformed to $|\alpha e^{ir_{0}\tau}\rangle$ after a time
$\tau$, based on the Poissonian distribution.  In the Wigner representation, each Fock
constituent has a rotation rate $r_{w,n}=r_{n}/n=r_0$ and the Wigner function rotates as a whole with rotation
angle $r_{0}\tau$, which is equal to the phase
shift measured in a resonator transmission measurement. This simple
behavior is most accurate in the far-detuned regime, where the dressed
resonator levels are still nearly equally spaced. At small detuning, where anharmonicity
becomes more significant, the various Fock components begin to rotate out of phase and
the Wigner function begins to display squeezing (see Fig. 3(a)).

To further illustrate the linear-nonlinear crossover
of the dispersive interaction, we investigated the Schr\"odinger cat state
$|\alpha=\sqrt{2}i\rangle+|\alpha=-\sqrt{2}i\rangle$. In Fig. 3(b) we display the reconstructed Wigner
functions for four different values of detuning $\Delta_f$, and three interaction times $\tau$. At
zero interaction time, the Wigner function contains two well-separated Gaussian peaks, arising from the two
coherent state elements, with clear interference fringes between the two peaks. The Wigner function
rotates counterclockwise with increasing time
$\tau$, as before. For detuning $\Delta_{f}/2\pi=-56\ \text{MHz}$, the shape of
the Wigner function is almost preserved as $\tau$
increases, consistent with expectations from the linear approximation.
As the absolute value of detuning $\Delta_{f}$ decreases, the rotation speed increases
and distortion becomes significant. The two Gaussian peaks
are twisted and the interference fringes deform
with time, due to
the differing rotation speeds of the Fock components. This strongly nonlinear
effect is most prominent for zero detuning, where the rotation rate scales as $r_{w,n}\propto1/\sqrt{n}$.\cite{ladder,nonlinear}

The examples above were measured with
a qubit initialized in its ground state, for which qubit relaxation and
phase decoherence are less important. We also examined the response of a resonator interacting
with a qubit initialized in the excited state. Figure 4(a) shows the Wigner function for a resonator initialized in $|0\rangle+|1\rangle$
after interacting with an excited-state qubit at detuning $\Delta_{f}/2\pi=-35\ \text{MHz}$. In contrast
to the response seen with a ground-state qubit, we observe the opposite rotation direction,
indicating a negative shift in photon frequency and phase. The nonclassical negative quasi-probabilities of the Wigner
function diminish much more quickly, in accordance with the rapid decoherence of the off-diagonal terms in the density matrix. In
addition, for three different final detunings $\Delta_f$ (Fig. 4(b)), the phase rotation angle of $\rho_{01}$
is no longer linearly dependent on $\tau$ but shows a periodic $S$-shaped
structure (Fig. 4(b)), which can be related to qubit relaxation.

\begin{figure}[H]
\begin{center}
\includegraphics[width=1.0\columnwidth,clip]{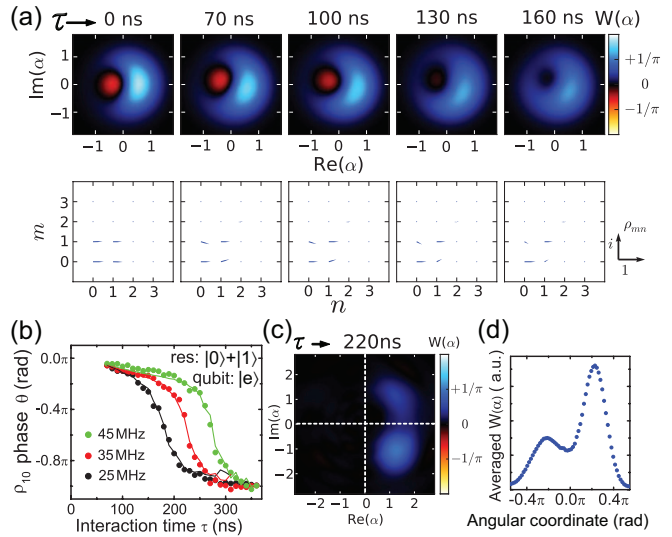}
\caption{~\label{Fig4}~(Color online)
Measurement of photon states after dispersive interaction with an
excited-state qubit. To minimize qubit relaxation during the interaction, we use a higher tuning ramp speed of $k=7\,\text{MHz}/\text{\, ns}$. The qubit was reset
to its ground state prior to the tomography measurement, by transferring
the excitation to a nearby two-level state using an $i$-swap gate.\cite{MatteoScience} (a) Wigner
functions and density matrices for the initial resonator  state $|0\rangle+|1\rangle$.
The dispersive interaction was measured for detuning $\Delta_{f}/2\pi=-35{\color{red}\ }\text{MHz}$
for five different durations $\tau$. (b) Rotation angle for $\rho_{10}$ vs. $\tau$ for three different detunings $\Delta_{f}$. Solid lines are numerical simulations of the Lindblad
master equation including both system decoherence and the third
level of the qubit. However, we have to use a shorter qubit decoherence time to yield reasonable agreement between the experimental
and theoretical results. For $|\Delta_{f}|/2\pi$ of $25\ \text{MHz}$, $35\ \text{MHz}$, and $45\ \text{MHz}$,
the qubit relaxation times $T_1$ were $200\ \text{ns}$, $220\ \text{ns}$, and $250\ \text{ns}$, respectively. (c) Wigner
function for a coherent state after interacting with an excited-state qubit
for $\tau=220\ \text{ns}$ at $\Delta_{f}/2\pi=-75\ \text{MHz}$. (d) Radially averaged $W(\alpha)$ as a function
of the angular coordinate, extracted from the Wigner function in (c). }
\end{center}
\end{figure}

To understand the effects of qubit relaxation, we investigated the response of a coherent photon state (Fig. 4(c)). During the dispersive
interaction, the qubit can relax at any moment, performing a stochastic quantum jump in its time trajectory. As a consequence, the
coherent peak in the Wigner function will initially rotate clockwise, but when the qubit relaxes,
switch to a counterclockwise rotation for the remainder of the dispersive interaction. The amplitude
of the jump probability is determined by the qubit relaxation time $T_1$ and
decays exponentially as time increases. The reconstructed
Wigner function thus comprises the main coherent peak (from the excited-state qubit) accompanied by a small peak with
a tail (due to rotation from the ground-state qubit). Numerical simulations of the Lindblad master
equation are quantitatively consistent with the experimental results (not shown).

For a general resonator state interacting with an excited qubit, the qubit relaxation and the mixed qubit state
lead to a decoherence effect on the final photon state. The decoherence effect is averaged over many individual quantum jumps
in the ensemble measurement required for collecting quantum statistics.
For a resonator initialized in $|0\rangle+|1\rangle$, the excited qubit
disperses the photons with a phase shift, i.e.,
rotates the horizontal off-diagonal vector $\rho_{01}$ to an angle $\theta(\tau)$.
The off-diagonal terms from qubit quantum jumps are
represented by vectors with rotation angles distributed over the range
$[\theta(\tau),-\theta(\tau)]$ and amplitudes
corresponding to the probability of the specific quantum jump.
Averaging over all the vectors resulting from quantum jumps decreases
the phase rotation angle of $\rho_{01}$ when $\theta$ is small, i.e. still an acute angle.
When $\theta$ is larger than $\pi/2$, the averaged vectors from the qubit
relaxation increase the phase rotation angle. This explains the
S-shape structure in the time-dependent phase rotation angles (Fig. 4(b)).
These data are in fact predicted (Fig. 4(b)) from numerical simulations using the Lindbald
master equation (cf. Appendix B) incorporating the third qubit level, qubit decoherence, and TLSs.

\section{CONCLUSIONS}

In conclusion, we have used an adiabatic dynamic control of the qubit to study its dispersive interaction with
a cavity resonator. In this dynamic scheme, the frequency shift of multi-photon
Fock states is extended to the strongly non-linear dispersive regime, close
to zero detuning. We can fully control and measure the accumulated phase shift of complex photon states,
demonstrated by the excellent agreement between the experimental data and the theoretical prediction.
In the strongly nonlinear dispersive regime, a quantum Kerr effect is observed for the coherent and other
non-classical photon states: We illustrate the nonlinear response of photons to the refractive index
effect of our artificial atom using Wigner tomography. Furthermore, we reveal the distinct phase shift of
photons induced by the excited state qubit, and interpret the peculiar photon decoherence resulting
from qubit state relaxation.

\section*{\uppercase{Acknowledgments}}

This work was supported by IARPA under ARO award W911NF-08-01-0336 and under ARO award W911NF-09-1-0375. M.M. acknowledges support from an Elings Postdoctoral Fellowship. Devices were made at the UC Santa Barbara Nanofabrication facility, a part of the NSF-funded National Nanotechnology Infrastructure Network.

\appendix
\section{EXPERIMENTAL SETUP}

The superconducting quantum circuit used three coplanar resonators
interconnected by two superconducting phase qubits,\cite{shellgame,haohua3,MatteoScience} although in this
experiment we used only one qubit and one resonator. The fabrication
process is a combination of photolithograhpy and
plasma etching on a multi-layered structure, with a superconducting rhenium
base film, Al/AlOx/Al Josephson junctions and amorphous silicon
for the shunt capacitor and wiring crossover dielectrics.\cite{John} An aluminum sample
box containing the wire-bonded device was mounted on the mixing
chamber of a dilution refrigerator operating at $\simeq25\ \text{mK}$.

The quasi-DC $\sigma_{z}$ tuning pulse was generated by custom electronics including an
FPGA-controlled digital-to-analog (DAC) converter.\cite{John} Two similar DAC
outputs, the I and Q signals of an IQ mixer, provided the sideband mixing,
phase, and pulse-shape control of the carrier signal from a microwave
generator to produce the $\sigma_{x,y}$ pulse. The qubit and resonator
frequencies were determined by spectroscopic measurements. The
qubit-resonator coupling strength $g$ and the resonator microwave pulse amplitude $\alpha$ were calibrated
using a photon population analysis for a coherent state.\cite{max1}
More details on the microwave calibration procedure are described in the
supplementary information of Ref. 19.

\section{MODELING AND SIMULATION}
The qubit-resonator
coupled system is described by the Jaynes-Cummings Hamiltonian,\cite{QuantumOptics}
 $H=\hbar\omega_{r}(a^{+}a+1/2)+\hbar\omega_{q}\sigma_{z}/2+\hbar g(\sigma^{+}a+a^{+}\sigma^{-})$,
where $\hbar\omega_{r}$ is the single photon energy of the relevant electromagnetic
field mode, $\hbar\omega_{q}$ the excitation energy of the qubit, and $a^{+}$($a$) and $\sigma^{+}$($\sigma^{-}$)
the creation (destruction) operators for the resonator photons and the qubit, respectively. With a coupling strength
$g$, the interaction term describes energy exchange between the qubit
and the resonator. The photon energy is effectively dispersed by
virtual energy exchange when an electromagnetic wave transmits through
the resonator at a detuning $\Delta=\omega_{q}-\omega_{r}$.
In the Jaynes-Cummings model,
the qubit-resonator interaction leads to two eigenenergies $n\hbar\omega_{r}\pm\frac{\hbar}{2}\sqrt{4g^{2}n+\Delta^{2}}$
for each state pair $|n,g\rangle$ and $|n-1,e\rangle$ ($\Delta<0$
is assumed). As shown in Fig. 1(b), the energy of a \emph{n}-photon
state is dispersed to

\begin{align}
ne_{p,g}= & \,\, E_{|n,g\rangle}-E_{|0,g\rangle}\nonumber \\
= & \,\, n\hbar\omega_{r}+\frac{\hbar}{2}(\sqrt{4g^{2}n+\Delta^{2}}-|\Delta|),\end{align}
when interacting with a ground state qubit. With the excited state
qubit, the dispersed energy becomes

\begin{align}
ne_{p,e}= & \,\, E_{|n,e\rangle}-E_{|0,e\rangle}\nonumber \\
= & \,\, n\hbar\omega_{r}-\frac{\hbar}{2}(\sqrt{4g^{2}(n+1)+\Delta^{2}}-\sqrt{4g^{2}+\Delta^{2}}).\end{align}
In the far-detuned regime, the linear approximation is applied and
the \emph{n}-photon energy is
simplified to be \begin{align}
ne_{p,\sigma_{z}}\simeq & \,\, n\hbar\omega_{r}+\sigma_{z}n\hbar g^{2}/\Delta\end{align}
for the ground-state and excited qubit (represented by $\sigma_{z}$).

The dispersive qubit-resonator interaction can be
strongly affected by the qubit relaxation and decoherence. Our numerical
simulations used the Markovian Lindblad master equation for system
density matrix $\rho$,

\[
\frac{d\rho}{dt}=-\frac{i}{\hbar}[H(t),\rho]+\sum_{L_{1},L_{2}}L\cdot\rho\cdot L^{\dagger}-\frac{1}{2}L^{\dagger}\cdot L\cdot\rho-\frac{1}{2}\rho\cdot L^{\dagger}\cdot L,\]
where $H(t)$ is the instantaneous total Hamiltonian and two superoperators,
$L_{1}=1/\sqrt{T_{1}}\sigma^{-}$ and $L_{2}=1/\sqrt{T_{2}}\sigma^{+}\sigma^{-}$,
describe qubit relaxation and decoherence. To simulate
the phase qubit, the third qubit level can be added in the
numerical master equation. We find that for the photon state phase shift
induced by a ground state qubit, the numerical results of the
master equation including decoherence and the third level contribution
are very close to the analytic solution of the Jaynes-Cummings Hamiltonian without dissipation, with
differences smaller than our experimental resolution. For photons
interacting with an excited qubit, in contrast, numerical simulations
with both decoherence and the third level have to be applied when comparing experimental data
with theoretical results (see Fig. 4(b)).

\bibliographystyle{apsrev}

\begin{thebibliography}{40}

\bibitem{QuantumOptics}D. E. Walls and G. J. Milburn, \emph{Quantum Optics} (Spring-Verlag, Berlin, Germany, 1994).

\bibitem{CavityQEDMabuchi}H. Mabuchi and A. C. Doherty, Science \textbf{298}, 1372 (2002).

\bibitem{CavityQEDBook}S. Haroche and J. -M. Raimond, \emph{Exploring the Quantum - Atoms, Cavities and Photons} (Oxford Univ. Press, Oxford, UK, 2006).

\bibitem{OpCavity}I. Schuster, A. Kubanek, A. Fuhrmanek, T. Puppe, P. W. H. Pinkse, K. Murr, and G. Rempe, Nat. Phys. \textbf{4}, 382 (2008).

\bibitem{QDotETH}K. Hennessy, A. Badolato, M. Winger, D. Gerace, M. Atat\"{u}re, S. Gulde, S. F\"{a}lt, E. L. Hu, and A. Imamo\u{g}lu, Nature (London) \textbf{445}, 896 (2007).

\bibitem{Horizons}R. J. Schoelkopf and S. M. Girvin, Nature (London) \textbf{451}, 664 (2008).

\bibitem{ScQubitReview}John Clarke and Frank K. Wilhelm, Nature (London) \textbf{453}, 1031 (2008).

\bibitem{QuaEnginneering}Y. Makhlin, G. Schon, and A. Shnirman, Rev. Mod. Phys. \textbf{73}, 357 (2001).

\bibitem{PhysToday}J. Q. You and Franco Nori, Phys. Today \textbf{58(11)}, 42 (2005).

\bibitem{Yale1}A. Wallraff, D. I. Schuster, A. Blais, L. Frunzio, R.- S. Huang, J. Majer, S. Kumar, S. M. Girvin, and R. J. Schoelkopf, Nature (London) \textbf{431}, 162 (2004).

\bibitem{ChiorescuNat04}I. Chiorescu, P. Bertet, K. Semba, Y. Nakamura, C. J. P. M. Harmans, and J. E. Mooij, Nature (London) \textbf{431}, 159 (2004).

\bibitem{max1}Max Hofheinz, E. M. Weig, M. Ansmann, Radoslaw C. Bialczak, Erik Lucero, M. Neeley, A. D. O'Connell, H. Wang, John M. Martinis, and A. N. Cleland, Nature (London) \textbf{454}, 310 (2008).

\bibitem{neccQED}O. Astafiev, K. Inomata, A. O. Niskanen, T. Yamamoto, Yu. A. Pashkin, Y. Nakamura, and J. S. Tsai, Nature (London) \textbf{449}, 588 (2007).

\bibitem{SimmondsBusNature}Mika A. Sillanpää, Jae I. Park, and Raymond W. Simmonds, Nature (London) \textbf{449}, 438 (2007).

\bibitem{IBM}R. H. Koch, G. A. Keefe, F. P. Milliken, J. R. Rozen, C. C. Tsuei, J. R. Kirtley, and D. P. DiVincenzo, Phys. Rev. Lett. \textbf{96}, 127001 (2006).

\bibitem{NTT}J. Johansson, S. Saito, T. Meno, H. Nakano, M. Ueda, K. Semba, and H. Takayanagi, Phys. Rev. Lett. \textbf{96}, 127006 (2006).

\bibitem{wallraffPRL05}A. Wallraff, D. I. Schuster, A. Blais, L. Frunzio, J. Majer, M. H. Devoret, S. M. Girvin, and R. J. Schoelkopf, Phys. Rev. Lett. \textbf{95}, 060501 (2005).

\bibitem{SchusterNat}D. I. Schuster, A. A. Houck, J. A. Schreier, A. Wallraff, J. M. Gambetta, A. Blais, L. Frunzio, J. Majer, B. Johnson, M. H. Devoret, S. M. Girvin, and R. J. Schoelkopf, Nature (London) \textbf{445}, 515 (2007).

\bibitem{max2}Max Hofheinz, H. Wang, M. Ansmann, Radoslaw C. Bialczak, Erik Lucero, M. Neeley, A. D. O'Connell, D. Sank, J. Wenner, John M. Martinis, and A. N. Cleland, Nature (London) \textbf{459}, 546 (2009).

\bibitem{ladder}J. M. Fink, M. Göppl, M. Baur, R. Bianchetti, P. J. Leek, A. Blais, and A. Wallraff, Nature (London) \textbf{454}, 315 (2008).

\bibitem{Deppe}Frank Deppe, Matteo Mariantoni, E. P. Menzel, A. Marx, S. Saito, K. Kakuyanagi, H. Tanaka, T. Meno, K. Semba, H. Takayanagi, E. Solano, and R. Gross, Nat. Phys. \textbf{4}, 686 (2008).

\bibitem{cavityRabi}M. Brune, F. Schmidt-Kaler, A. Maali, J. Dreyer, E. Hagley, J. M. Raimond, and S. Haroche, Phys. Rev. Lett. \textbf{76}, 1800 (1996).

\bibitem{nonlinear}Lev S. Bishop, J. M. Chow, Jens Koch, A. A. Houck, M. H. Devoret, E. Thuneberg, S. M. Girvin, and R. J. Schoelkopf, Nat. Phys. \textbf{5}, 105 (2009).

\bibitem{Reed}M. D. Reed, L. DiCarlo, B. R. Johnson, L. Sun, D. I. Schuster, L. Frunzio, and R. J. Schoelkopf, Phys. Rev. Lett. \textbf{105}, 173601 (2010).

\bibitem{John}John M. Martinis, Quantum Information Processing \textbf{8}, 81 (2009).

\bibitem{ErikPRL}Erik Lucero, M. Hofheinz, M. Ansmann, Radoslaw C. Bialczak, N. Katz, Matthew Neeley, A. D. O'Connell, H. Wang, A. N. Cleland, and John M. Martinis, Phys. Rev. Lett. \textbf{100}, 247001 (2008).

\bibitem{ErikPRA}Erik Lucero, Julian Kelly, Radoslaw C. Bialczak, Mike Lenander, Matteo Mariantoni, Matthew Neeley, A. D. O'Connell, Daniel Sank, H. Wang, Martin Weides, James Wenner, Tsuyoshi Yamamoto, A. N. Cleland, and John M. Martinis, Phys. Rev. A \textbf{82}, 042339 (2010).

\bibitem{MatthewPRB}Matthew Neeley, M. Ansmann, Radoslaw C. Bialczak, M. Hofheinz, N. Katz, Erik Lucero, A. O'Connell, H. Wang, A. N. Cleland, and John M. Martinis, Phys. Rev. B \textbf{77}, 180508 (2008)

\bibitem{photonQND}B. R. Johnson, M. D. Reed, A. A. Houck, D. I. Schuster, Lev S. Bishop, E. Ginossar, J. M. Gambetta, L. DiCarlo, L. Frunzio, S. M. Girvin, and R. J. Schoelkopf, Nat. Phys. \textbf{6}, 663 (2010).

\bibitem{berry}M. V. Berry, Proc.R. Soc. Lond. A Math. Phys. Sci. \textbf{392}, 45 (1984).

\bibitem{cavityLamb}M. Brune, P. Nussenzveig, F. Schmidt-Kaler, F. Bernardot, A. Maali, J. M. Raimond, and S. Haroche, Phys. Rev. Lett. \textbf{72}, 3339 (1994).

\bibitem{cavityQED}Sébastien Gleyzes, S. Kuhr, C. Guerlin, J. Bernu, S. Deléglise, Ulrich Busk Hoff, M. Brune, Jean-Michel Raimond, and S. Haroche, Nature (London) \textbf{446}, 297 (2007).

\bibitem{cavityphotonQND}Christine Guerlin, J. Bernu, S. Deléglise, C. Sayrin, S. Gleyzes, S. Kuhr, M. Brune, Jean-Michel Raimond, and S. Haroche, Nature (London) \textbf{448}, 889 (2007).

\bibitem{qudit}Matthew Neeley, M. Ansmann, Radoslaw C. Bialczak, M. Hofheinz, Erik Lucero, A. O'Connell, Daniel Sank, H. Wang, J. Wenner, A. N. Cleland, M. R. Geller, and John M. Martinis, Science \textbf{325}, 722 (2009).

\bibitem{haohua2}H. Wang, M. Hofheinz, M. Ansmann, R. C. Bialczak, Erik Lucero, M. Neeley, A. D. O'Connell, D. Sank, M. Weides, J. Wenner, A. N. Cleland, and John M. Martinis, Phys. Rev. Lett. \textbf{103}, 200404 (2009).

\bibitem{WallraffLamb}A. Fragner, M. G\"{o}eppl, J. M. Fink, M. Baur, R. Bianchetti, P. J. Leek, A. Blais, and A. Wallraff, Science \textbf{322}, 1357 (2008).

\bibitem{shellgame}Matteo Mariantoni, H. Wang, Radoslaw C. Bialczak, M. Lenander, Erik Lucero, M. Neeley, A. D. O'Connell, D. Sank, M. Weides, J. Wenner, T. Yamamoto, Y. Yin, J. Zhao, John M. Martinis, and A. N. Cleland, Nat. Phys. \textbf{7}, 287 (2011).

\bibitem{haohua3}H. Wang, M. Mariantoni, Radoslaw C. Bialczak, M. Lenander, Erik Lucero, M. Neeley, A. D. O'Connell, D. Sank, M. Weides, J. Wenner, T. Yamamoto, Y. Yin, J. Zhao, John M. Martinis, and A. N. Cleland, Phys. Rev. Lett. \textbf{106}, 060401 (2011).

\bibitem{MatteoScience}Matteo Mariantoni, H. Wang, T. Yamamoto, M. Neeley, Radoslaw C. Bialczak, Y. Chen, M. Lenander, Erik Lucero, A. D. O'Connell, D. Sank, M. Weides, J. Wenner, Y. Yin, J. Zhao, A. N. Korotkov, A. N. Cleland, and John M. Martinis, Science \textbf{334}, 61 (2011).


\end{thebibliography}


\end{document}